\documentclass[secnumarabic, nobibnotes, aps,prl,twocolumn,superscriptaddress]{revtex4-2} 
\usepackage{amsmath,graphicx,subcaption,array}
\usepackage{dcolumn,soul}%

\usepackage{amsthm}
\usepackage[figuresright]{rotating}%
\usepackage{algorithm, algorithmicx, algpseudocode}
\usepackage{listings}%
\usepackage{hyperref}
\usepackage{siunitx}
\usepackage{mhchem}

\makeatletter
\def\uns{\ifmmode\,\else$\,$\fi}%

\makeatother

\received{XX XX 2025}
\revised{XX XX 2025}
\accepted{XX XX 2025}

\begin{document}

\title{First systematic experimental 2D mapping of linearly polarized $\gamma$-ray polarimetric distribution in relativistic Compton scattering}

\author{Kaijie Chen}
\affiliation{ShanghaiTech University, Shanghai 201210, China}
\affiliation{Shanghai Advanced Research Institute, Chinese Academy of Sciences, Shanghai 201210, China}

\author{Xiangfei Wang}
\affiliation{Shanghai Institute of Applied Physics, Chinese Academy of Sciences, Shanghai 201800, China}
\affiliation{University of Chinese Academy of Sciences, Beijing 100049, China}

\author{Hanghua Xu}
\email{xuhh@sari.ac.cn}
\affiliation{Shanghai Advanced Research Institute, Chinese Academy of Sciences, Shanghai 201210, China}
\affiliation{University of Chinese Academy of Sciences, Beijing 100049, China}

\author{Gongtao Fan}
\email{fangt@sari.ac.cn}
\affiliation{Shanghai Advanced Research Institute, Chinese Academy of Sciences, Shanghai 201210, China}
\affiliation{Shanghai Institute of Applied Physics, Chinese Academy of Sciences, Shanghai 201800, China}
\affiliation{University of Chinese Academy of Sciences, Beijing 100049, China}

\author{Zirui Hao}
\affiliation{Shanghai Advanced Research Institute, Chinese Academy of Sciences, Shanghai 201210, China}

\author{Longxiang Liu}
\affiliation{Shanghai Advanced Research Institute, Chinese Academy of Sciences, Shanghai 201210, China}

\author{Yue Zhang}
\affiliation{Shanghai Advanced Research Institute, Chinese Academy of Sciences, Shanghai 201210, China}

\author{Sheng Jin}
\affiliation{Shanghai Institute of Applied Physics, Chinese Academy of Sciences, Shanghai 201800, China}
\affiliation{University of Chinese Academy of Sciences, Beijing 100049, China}

\author{Zhicai Li}
\affiliation{Shanghai Advanced Research Institute, Chinese Academy of Sciences, Shanghai 201210, China}
\affiliation{School of Nuclear Science and Technology, University of South China, Hengyang 421001, China}

\author{Pu Jiao}
\affiliation{Shanghai Advanced Research Institute, Chinese Academy of Sciences, Shanghai 201210, China}
\affiliation{College of Physics, Henan Normal University, Xinxiang 453007, China}

\author{Qiankun Sun}
\affiliation{Shanghai Institute of Applied Physics, Chinese Academy of Sciences, Shanghai 201800, China}
\affiliation{University of Chinese Academy of Sciences, Beijing 100049, China}

\author{Zhenwei Wang}
\affiliation{Shanghai Institute of Applied Physics, Chinese Academy of Sciences, Shanghai 201800, China}
\affiliation{University of Chinese Academy of Sciences, Beijing 100049, China}

\author{Mengdie Zhou}
\affiliation{Shanghai Advanced Research Institute, Chinese Academy of Sciences, Shanghai 201210, China}
\affiliation{College of Physics, Henan Normal University, Xinxiang 453007, China}

\author{Mengke Xu}
\affiliation{Shanghai Institute of Applied Physics, Chinese Academy of Sciences, Shanghai 201800, China}
\affiliation{University of Chinese Academy of Sciences, Beijing 100049, China}

\author{Hongwei Wang}
\affiliation{Shanghai Advanced Research Institute, Chinese Academy of Sciences, Shanghai 201210, China}
\affiliation{Shanghai Institute of Applied Physics, Chinese Academy of Sciences, Shanghai 201800, China}
\affiliation{University of Chinese Academy of Sciences, Beijing 100049, China}

\author{Wenqing Shen}
\affiliation{Shanghai Advanced Research Institute, Chinese Academy of Sciences, Shanghai 201210, China}
\affiliation{University of Chinese Academy of Sciences, Beijing 100049, China}

\author{Yugang Ma}
\email{mayugang@fudan.edu.cn}
\affiliation{Institute of Modern Physics, Fudan University, Shanghai 200433, China}
\affiliation{ShanghaiTech University, Shanghai 201210, China}
\date{\today}

\newpage
\begin{abstract}
The interaction of photons with relativistic electrons constitutes a fundamental electromagnetic process whose polarization transfer mechanics remain incompletely characterized. We report the first systematic measurement of spatial polarization distribution for $\gamma$-rays generated via \SI{45}{\degree} slant inverse Compton scattering (ICS) between linearly polarized \SI{0.117}{\eV} photons and \SI{3.5}{\GeV} electrons, performing full 2D mapping of intensity, polarization angle (AOP), and degree of polarization (DOP). Measurements reveal an asymmetric beam profile along the laser's polarization direction that resembles \SI{180}{\degree} backward ICS observations. The central beam region exhibits DOP $\approx$ 1.0 with AOP rigidly aligned at \SI{45}{\degree}, while peripheral regions display complex non-uniform polarization distributions. These findings confirm quantum electrodynamics predictions of near-complete polarization transfer along the beam axis in slant geometries, thus establishing slant scattering as a viable alternative to head-on configurations for generating high DOP $\gamma$-rays. 
\end{abstract}

\maketitle

\section{INTRODUCTION}\label{sec::introduction}
The interaction of photons with electrons is one of the most fundamental electromagnetic processes and has been systematically studied over a long history since the 1920s~\cite{comptonQuantum1923,comptonWavelength1924}. Within the theoretical framework of quantum electrodynamics (QED), the scattered photons naturally inherit the polarization information of incident photons. And if the incident electrons are relativistic, in other words, endowed a high momentum, then the scattered photons will propagate predominantly along the direction of the electron's momentum~\cite{choiFinalphoton1987,landauClassicalTheoryFields1980, jacksonClassicalElectrodynamics1998, Greiner_QED}. 

The properties of the photons scattered from the electrons can be well described by QED~\cite{Greiner_QED}. Especially in the non-relativistic case, employing the Stokes parameters $\xi_i$, the spatial distribution of the degree of polarization (DOP) and angle of polarization (AOP) can be obtained~\cite{mcmasterMatrixRepresentationPolarization1961,hamzawyCompton2016}. For instance, in the interaction between the complete linearly polarized incident photons ($\xi_2$=0, $\xi_1^2 + \xi_3^2$=1) and non-relativistic electrons, QED predicts that the scattered photons remain linearly polarized, with the spatial intensity distribution exhibiting azimuthal asymmetry~\cite{Greiner_QED,landauClassicalTheoryFields1980}. Furthermore, the non-relativistic case of polarized photons scattering with electrons, Compton scattering, has been experimentally validated with high precision and successfully applied to Compton camera technology~\cite{goDemonstrationNuclearGammaray2024} and astrophysics~\cite{Science_polarizedG_Astrophysics, Science_polarizedG_CygnusX1,2022_taverna_Science}.

However, in relativistic regimes, QED-based descriptions of inverse Compton scattered photon properties necessitate sophisticated treatments. Approximations remain nearly unavoidable during theoretical derivations or calculations~\cite{liPolarized2020,SunCHangchun_HIGS,choiFinalphoton1987}. Recently, multiple research groups have simulated polarization distributions of scattered photons from polarized photon and relativistic electron beam collisions using Monte Carlo methods within QED or classical electrodynamics frameworks, yielding consistent results~\cite{maCompactPolarizedXRay2023,zhangExperimentalDemonstrationPolarization2024,Tsinghua_Zhang, liPolarized2020, SunCHangchun_HIGS, Filipescu_Monte_Carlo, ChiZhijun_Thomson, ELI-NP_Polarization}. 

The experimental investigations into the polarization properties of inverse Compton scattering have also made some progress over the pass two decades. In 2007, the spatial intensity distribution of scattered $\gamma$-rays, produced from the \SI{1.17}{\eV} ($\lambda$ = \SI{1.064}{\um}) linearly polarized laser photons \SI{180}{\degree} backward scattering on \SI{1.0}{\GeV} electrons at NewSUBARU was recorded by imaging plates~\cite{NewSUBARU}. In 2011 the spatial intensity distribution of the scattered $\gamma$-rays produced from the \SI{3.29}{\eV} ($\lambda$ = \SI{378}{\nm}) linearly polarized Free Electron Laser (FEL) \SI{180}{\degree} backward scattering on \SI{680}{\MeV} electrons at HI$\gamma$S was measured with a $\gamma$ imaging system~\cite{SunCHangchun_HIGS}. Then in 2019, also at HI$\gamma$S, the DOP and AOP of the center of scattered $\gamma$-ray beam in the \SI{180}{\degree} backward scattering between linearly polarized \unit{\eV} FEL and \unit{\GeV} electrons were extracted by a Time-Of-Flight (TOF) array~\cite{HIGS_Yan_Nature_Photonics}. Similarly, at UVSOR-III, two experiments were conducted to measure the spatial polarizations distribution of polarized $\gamma$-rays generated by a \SI{90}{\degree} collision between a \SI{750}{\MeV} electron beam and \SI{1.55}{\eV} ($\lambda$ = \SI{800}{\nm}) laser: in 2023, a circularly polarized laser was used, with measurements performed along the radius of $\gamma$ spot by HPGe, \ce{LaBr_{3}} detectors and magnetized iron~\cite{UVSOR_Taira_circularly_polarized}; in 2024, several axially symmetric polarized and optical vertex lasers were used, with measurements performed by only an image sensor~\cite{UVSOR_Taira_vortex}. Also in 2024, Tsinghua University experimentally demonstrated the average DOP in the central region of Thomson scattering X-ray~\cite{zhangExperimentalDemonstrationPolarization2024}. Notably, despite these experimental advancements, experimental data regarding the spatial distribution of DOP and AOP of the scattered photons from the relativistic electrons are not systematic and sufficient. The complete 2D polarization mapping within $\gamma$-ray beam cross-section has not to be experimentally realized yet. 

The scarcity of experimental data on the polarization properties of photons scattering from relativistic electrons hinders both theoretical and experimental research. In the absence of systematic data on the DOP and AOP distribution of the scattered photons, the above-mentioned theoretical models of photon and relativistic electron interaction cannot be verified. In those specialized experimental scenario involving the Inverse Compton Scattering (ICS) light source, such data are even more indispensable~\cite{Pietralla_EPJA, Pietralla_HIGS,beck02020,beck22017,bellerConstraint2013,kohriDifferential2018,zegersExtraction2007,zegersBeampolarization2003}, which directly determines the accuracy of polarization-dependent measurements~\cite{faggPolarizationMeasurementsNuclear1959, morehPolarizationMeasurementsResonantly1972}. Here, we report the first systematic measurement of the spatial polarization distribution of medium-energy (approximately \SI{3}{\MeV}) $\gamma$-ray photons produced from a \SI{45}{\degree} slant ICS between \SI{0.117}{\eV} ($\lambda$ = \SI{10.64}{\um}) \ce{CO_2} laser photons and \SI{3.5}{\GeV} relativistic electrons in the Shanghai Synchrotron Radiation Facility (SSRF) storage ring.

\section{RESULTS AND DISCUSSION}\label{sec::results}

\subsection{Experiment setup and method}\label{subsec::experiment_setup}
\begin{figure*}[htb]
    \includegraphics[width=\textwidth]{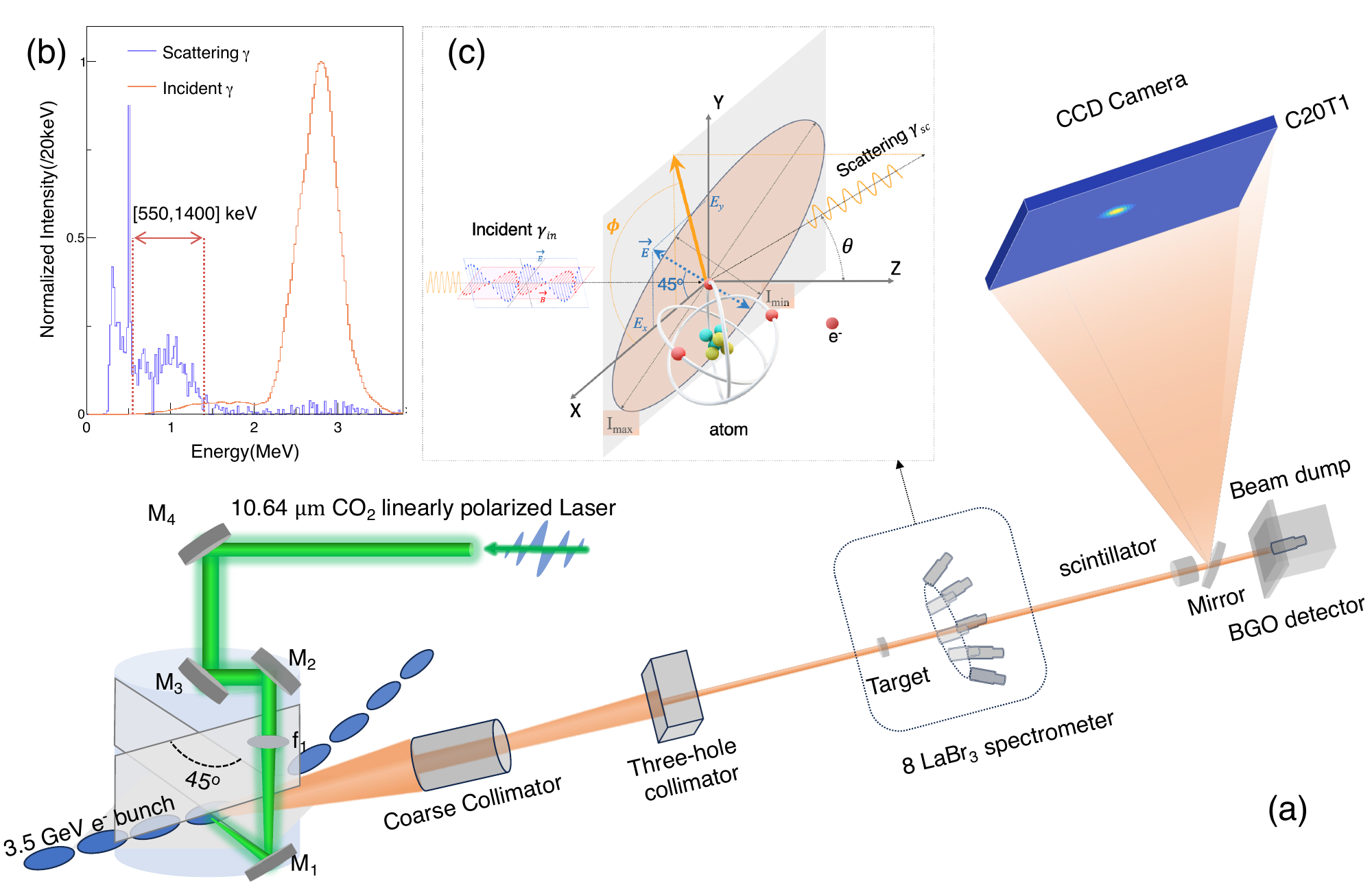}
    \caption{(a), the experiment setup, where the Compton process between the incident $\gamma$-ray and Ta target is represents in the inset (c). The spectra of incident $\gamma$-rays and secondary scattering $\gamma$-rays from Ta target are shown in the inset (b) at centre of $\gamma$-rays beam spot.}
    \label{fig::Setup}
\end{figure*}

The experiment was carried out on the Shanghai Laser Electron Gamma Source (SLEGS) beamline of SSRF~\cite{SLEGS_WangHW_Commissioning, Liu_LX_SLEGS_beamline}. The experiment setup is shown in Fig.~\ref{fig::Setup}. Focused linearly polarized \ce{CO_2} laser photons \SI{0.117}{\eV}  are aligned to collide with the \SI{3.5}{\GeV} electrons at \SI{45}{\degree}. During the experiment, the SSRF storage ring was operated in Top-Up mode with \SI{200}{\mA} electron beam current, while the \ce{CO_2} laser ran in gated continuous wave mode using an RF Enable signal, producing pulses with a \SI{1}{\kHz} repetition frequency and \SI{50}{\us} width (see S1). The DOP of the incident laser was carefully measured to be $>$ 93\% (see S2).

The spatial profile and intensity distribution of the $\gamma$-ray spot were recorded at 35 meters downstream from the collision point using a $\gamma$-ray imaging system~\cite{SLEGS_GMS}. 

To measure the transverse distribution of the AOP and DOP of the generated $\gamma$-ray beam, a $\gamma$-ray polarimeter based on Compton scattering, referred to as linearly polarized $\gamma$-ray scattering with a Tantalum target, was developed at the SLEGS. This method has been widely adopted and verified in Compton camera applications~\cite{goDemonstrationNuclearGammaray2024} and Compton polarimeter~\cite{1976_beck_PhysicalReviewC,1975_williams_PhysicalReviewC,1973_hardy_PhysicalReviewC,1995_vonderwerth_NuclearInstrumentsandMethods,2015_bizzeti_TheEuropeanPhysicalJournalA}. The geometry of Compton scattering between linearly polarized $\gamma$-ray and electrons of Ta target in a laboratory frame is shown in Fig.~\ref{fig::Setup} (c). The differential cross-section for Compton scattering of linearly polarized $\gamma$-rays is described by Klein–Nishina formula
\begin{equation}
    \frac{\mathrm{d}\sigma}{\mathrm{d}\Omega}(\theta,\phi,E_{\mathrm{in}})=\frac{r_{e}^2}{2}(\frac{E_{\mathrm{sc}}}{E_{\mathrm{in}}})^2(\frac{E_{\mathrm{sc}}}{E_{\mathrm{in}}}+\frac{E_{\mathrm{in}}}{E_{\mathrm{sc}}}-\sin^2\Theta )
    \label{eq::Klein_Nishina_formula_polarization}
\end{equation}
where $r_{e}$ is the classical electron radius, $E_{\mathrm{in}}$ and $E_{\mathrm{sc}}$ are the energy of incident and scattered photons, respectively, $\Theta$ is the angle between the two polarization vectors. In terms of the $\theta$, the polar scattering angle, and $\phi$, the angle between the polarization plane of the incident $\gamma$-rays and the Compton scattering plane, respectively, the Eq.~\ref{eq::Klein_Nishina_formula_polarization} can be rewritten as

\begin{equation}
    \begin{split}
        \frac{\mathrm{d}\sigma}{\mathrm{d}\Omega}(\theta,\phi,E_{\mathrm{in}}) 
        & =\frac{r_{e}^2}{2}
        \left( \frac{E_{\mathrm{sc}}}{E_{\mathrm{in}}} \right)^2 \\
        &\cdot \left( 
            \frac{E_{\mathrm{sc}}}{E_{\mathrm{in}}} 
            + \frac{E_{\mathrm{in}}}{E_{\mathrm{sc}}} 
            - 2\sin^2\theta\cos^2\phi 
        \right)
    \end{split}
    \label{eq::Klein_Nishina_formula}
\end{equation}
The energies $E_{\mathrm{in}}$ and $E_{\mathrm{sc}}$ are related by expression
\begin{equation}
    E_{\mathrm{sc}}=\frac{E_{\mathrm{in}}}{1+\frac{E_{\mathrm{in}}}{m_{e}c^2}(1-\cos\theta)}
    \label{eq::energy_relationship}
\end{equation}
The Eq.~\ref{eq::Klein_Nishina_formula} shows that the azimuthal distribution of the secondary Compton photons is $\mathrm{cos^2\phi}$-like shape. The differential cross-section reaches its extreme values at $\phi=\SI{0}{\degree}$ and \SI{90}{\degree}. Its ratio is called Peak-to-Valley Ratio (PVR)
\begin{equation}
    \mathrm{PVR} = \frac{\frac{\mathrm{d}\sigma}{\mathrm{d}\Omega}(\theta,\SI{90}{\degree},E_{\mathrm{in}})}{\frac{\mathrm{d}\sigma}{\mathrm{d}\Omega}(\theta,\SI{0}{\degree},E_{\mathrm{in}})}
    \label{eq::PVR}
\end{equation}
The asymmetry which is proportional to the DOP of the incident $\gamma$-rays is defined as
\begin{equation}
    A(\theta,E_{\mathrm{in}}) =\frac{ \frac{\mathrm{d}\sigma}{\mathrm{d}\Omega}(\theta,\SI{90}{\degree},E_{\mathrm{in}})- \frac{\mathrm{d}\sigma}{\mathrm{d}\Omega}(\theta,\SI{0}{\degree},E_{\mathrm{in}})}{ \frac{\mathrm{d}\sigma}{\mathrm{d}\Omega}(\theta,\SI{90}{\degree},E_{\mathrm{in}})+ \frac{\mathrm{d}\sigma}{\mathrm{d}\Omega}(\theta,\SI{0}{\degree},E_{\mathrm{in}})}
    \label{eq::asymmetry}
\end{equation}
and can also be derived from the PVR. These results demonstrate a strong correlation between the asymmetry and both the detector's geometric angle $\theta$ and the incident $\gamma$-ray energy $E_{\mathrm{in}}$~\cite{mcmasterMatrixRepresentationPolarization1961,faggPolarizationMeasurementsNuclear1959, morehPolarizationMeasurementsResonantly1972}. Specifically, as the energy of $\gamma$-rays increases, the optimal geometric angle of detector $\theta_\mathrm{{max}}$ and the asymmetry will decrease. According to theoretical calculations, a monoenergetic $\gamma$-ray beam ($<$ \SI{4}{\MeV}) with full linear polarization generates substantial asymmetry in the transverse plane relative to beam propagation. Such pronounced asymmetry facilitates precise quantification of the $\gamma$-ray polarization state.

To find the theoretically expected asymmetry $A_{\mathrm{th}}$, a Monte Carlo
simulation was performed using GEANT4~\cite{agostinelliGEANT4aSimulationToolkit2003}, incorporating real beam parameters along with other inputs such as the target and detectors' dimensions and positions (see S3). In the polarization experiment, the differential cross-section can be represented by the azimuthal distribution of counts and the fitting of the function 
\begin{equation}
    N(\phi)=P_{0}-P_{1}\cos^2\frac{\pi(\phi-P_{2})}{180}
    \label{eq::polarization_fitting}
\end{equation}
where ${P_0}$ and $P_1$  are used to calculate the experimental asymmetries $A_{\mathrm{exp}}$ and their uncertainties, and $P_2$ is the phase shift (in degrees) associated with the AOP.
At last, the DOP of the incident $\gamma$-rays is obtained from
\begin{equation}
    \mathrm{DOP} = \frac{A_{\mathrm{exp}}}{A_{\mathrm{th}}}=\frac{P_{1}}{(2P_{0}-P_{1})A_{\mathrm{th}}}
    \label{eq::polarization_degree}
\end{equation}.

A precisely controlled \SI{1}{\mm} aperture collimator is employed to scan the transverse 2D distribution of polarized $\gamma$-rays. At each scanning position, the collimated $\gamma$-rays irradiate the cylindrical, purity of 99.9\% \ce{Ta} target with a diameter of \SI{6}{\mm} and thickness of \SI{10}{\mm}. The scattered secondary photons from the Ta target are then measured by an array of 8 \ce{LaBr_{3}(Ce)} detectors. These detectors are arranged azimuthally symmetric around the Ta target at a radial distance of \SI{22}{\cm}. The array geometry is shown in Fig.~\ref{fig::Setup} (a). 

Before the azimuthal distribution measurement, the collimator positions were scanned to determine the center point of the $\gamma$ beam spot. Combined with the $\gamma$-ray spot data obtained from the $\gamma$ imaging system, 49 linearly distributed scanning points were positioned symmetrically along four radial directions separated by \SI{45}{\degree}, as illustrated in Fig.~\ref{fig::angle_distribution}. At every scanning point, the incident polarized $\gamma$-rays' flux and energy were measured by the BGO detector placed in the beam dump (the details of the measurement are expounded in S4).

\subsection{Intensity azimuthal distribution}\label{subsec::intensity_azimuthal_distribution}
\begin{figure*}[htb]
\centering
     \begin{subfigure}{0.46\textwidth}
    \includegraphics[width=\textwidth]{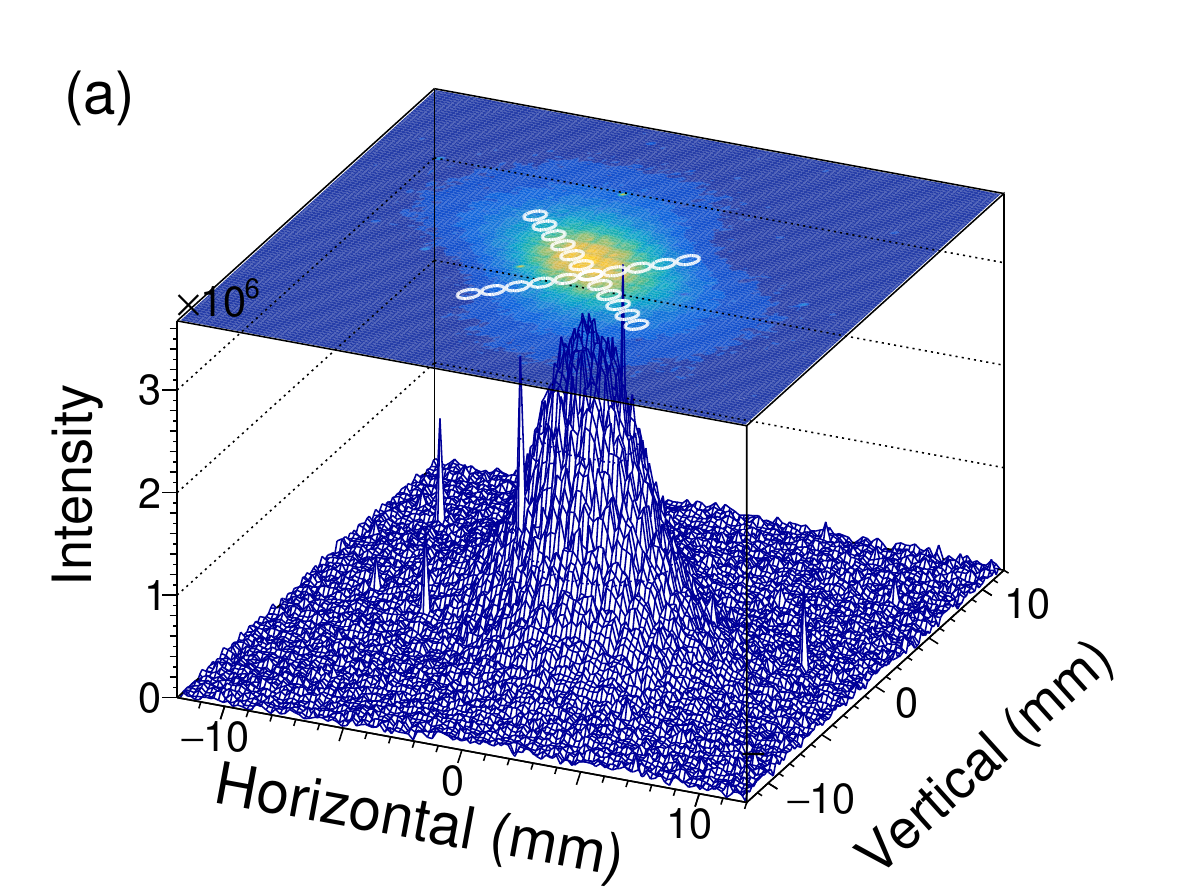}
    \end{subfigure}
    \begin{subfigure}{0.46\textwidth}
    \includegraphics[width=\textwidth]{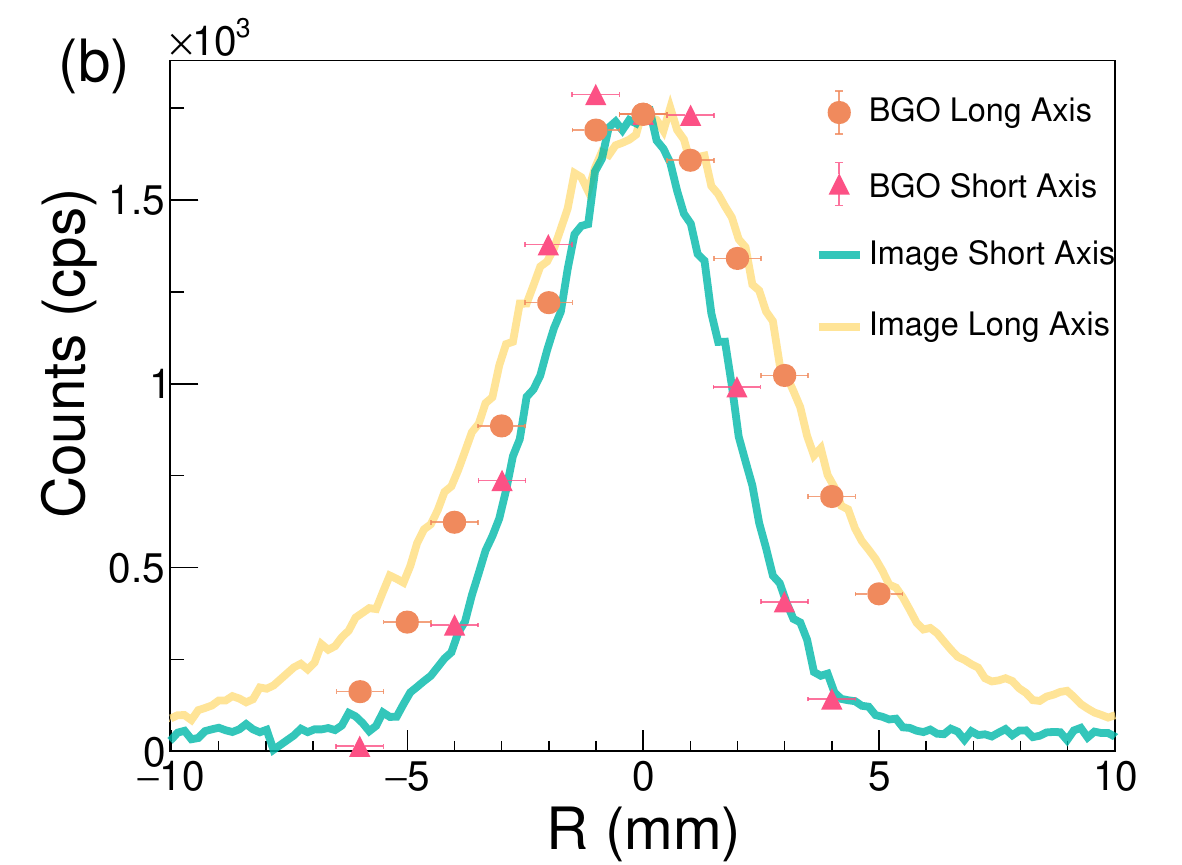}
    \end{subfigure}
\caption{(a) The measured spatial distributions of scattered polarized $\gamma$-rays intensity using the $\gamma$-ray imaging system and collimator scanning. 3D spatial distributions result is from the $\gamma$-ray imaging. And data points are from collimator scanning. (b) Comparison of the scattered polarized $\gamma$-ray intensities along the long and short axes between the $\gamma$-ray imaging system and collimator scanning.}
    \label{fig::intensity}
\end{figure*}

Figure.~\ref{fig::intensity} (a) shows the measured spatial distributions of incident $\gamma$ rays using the $\gamma$-ray image system. The collimator scanning points corresponding to the BGO counts in fig.~\ref{fig::intensity} (b) is draw in the top projection. It clearly shows that the intensity distribution of $\gamma$-rays generated from a linearly polarized laser photons and relativistic electrons is asymmetric, and is ``pinched" along the AOP of the laser beam. Our findings of spatial intensity distribution of the generated $\gamma$-ray under \SI{45}{\degree} slant ICS is similar with prior investigations at HI$\gamma$S~\cite{HIGS_Yan_Nature_Photonics} and NewSUBARU~\cite{NewSUBARU} under \SI{180}{\degree} backward ICS, and UVSOR-III under \SI{90}{\degree} slant ICS~\cite{UVSOR_Taira_vortex}. 

The $\gamma$ imaging system relys on scintillator fluorescence captured by CCD camera. Whereas the scattering effect of optical system and the $\gamma$ energy-dependent light yield of scintillator complicates precise measurement of $\gamma$-ray spot intensity distribution~\cite{2024_christy_OpticalMaterials,chewpraditkulScintillationPropertiesLuAG2009,maoOpticalScintillationProperties2008}, BGO detector can provide a direct measurement of the collimated $\gamma$-ray flux. The measurement of $\gamma$ intensity using BGO detector has been verified through a large number of experiments which features high reliability and accuracy~\cite{LiuLX_BGO_Unfold, ZhouMD_Co59_PRC,2025_sun_NST,2025_YangYX_NST,2025_jiao_NST}. 

Comparative analysis of normalized intensity distributions (BGO detector vs. $\gamma$-imaging; full dataset is shown in S5) reveals that conventional imaging methods produce only qualitative shape profiles for polarized $\gamma$-rays, whereas BGO scanning enables quantitative intensity mapping. Fig.~\ref{fig::intensity} (b) quantifies this disparity along beam-spot axes. To our knowledge, this represents the first precise measurement of linearly polarized $\gamma$-ray intensity distributions via scanned BGO detection. 

\subsection{AOP distribution}\label{subsec::polarization_direction_distribution}
\begin{figure*}[htb]
    \centering
    \includegraphics[width=\textwidth]{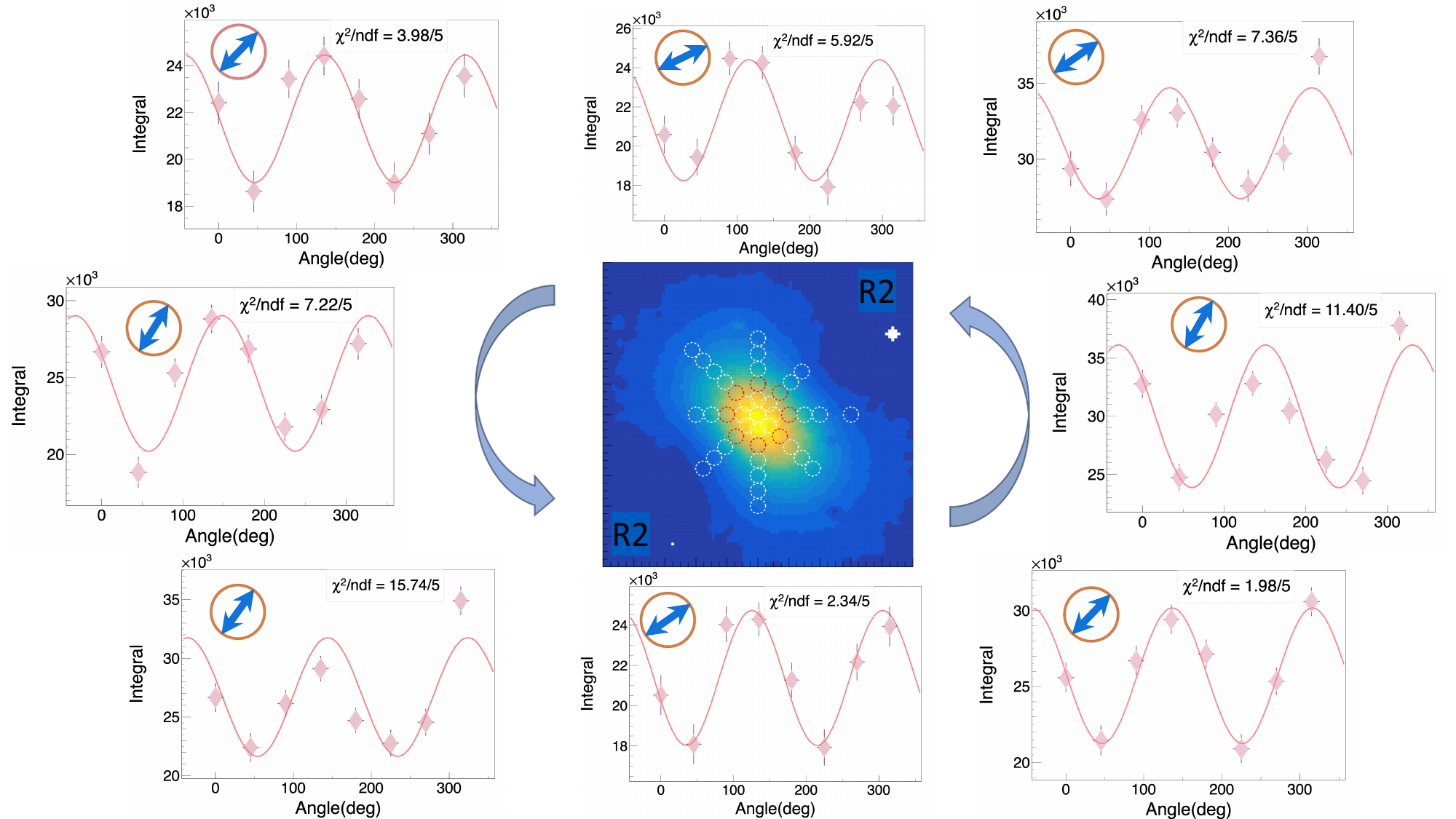}
    \caption{The profile of $\gamma$-rays without collimator is visualized in the center, where the all scanning points are presented by the white circles at the top projection and the points corresponding to the surround azimuthal distributions are highlighted with the red circles. The surround azimuthal distribution of the secondary scattered photons from Ta target when the $\phi$\SI{1}{\mm} collimator is aligned the position \SI{2}{\mm} away from the center of the $\gamma$-rays beam at the different directions.}
    \label{fig::angle_distribution}
\end{figure*}

\begin{figure*}[htb]
    \includegraphics[width=\textwidth]{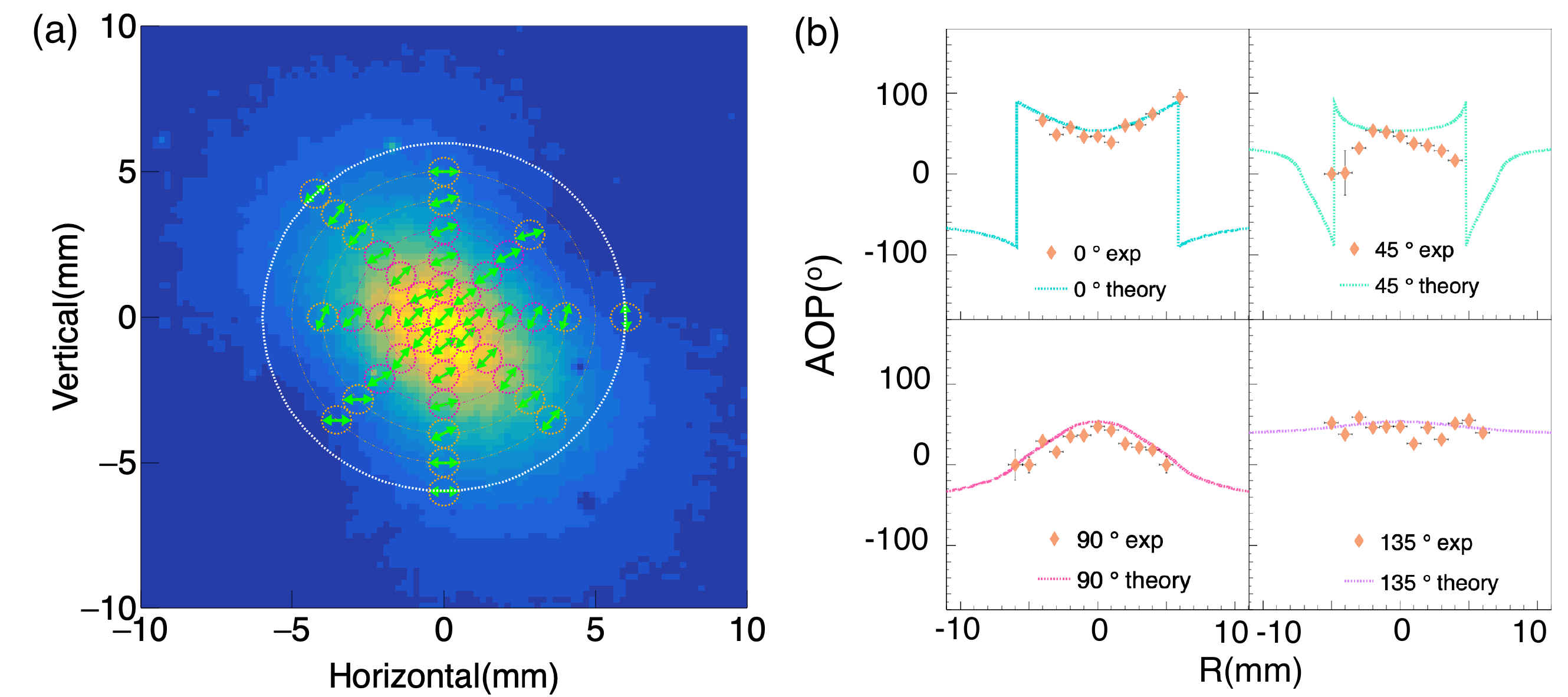}
    \caption{(a) the polarization direction (AOP) for each measurement point, where the circle with pink represents the $1/\gamma$.  A comparison between the experimental and calculation based on QED~\cite{WangXF_SLEGS_Polarization} is presented in (b).}
    \label{fig::Polar_Direction}
\end{figure*}

The \ce{LaBr_{3}(Ce)} array measured the azimuthal distributions of secondary scattered photons from the Ta target at precisely adjusted collimator scanning positions. A typical energy spectrum of the scattered photons measured by a \ce{LaBr_{3}(Ce)} detector is shown in Fig.~\ref{fig::Setup} (b). The fine geometric corrections of based on Monte Carlo simulations are implemented for each \ce{LaBr_{3}(Ce)} detector at every measurement point. The uncertainties come from the fluctuations of intrinsic efficiency, statistical errors and the fine geometric correction errors (detailed data analysis can be found in S5).

Figure~\ref{fig::angle_distribution} shows representative azimuthal distributions of secondary scattered photons from the Ta target measured by \ce{LaBr_{3}(Ce)} array. These were measured with the \SI{1}{\mm} diameter collimator positioned precisely \SI{2}{\mm} away from the incident $\gamma$-ray beam center along four radial directions. The red curves represent the fits using Equation~\ref{eq::polarization_fitting}. Among 28 experimental datasets, $\mathrm{\chi^2}$ $\le$ 10 for all eight measurement configurations, confirming that at 2 mm from the polarized $\gamma$-ray beam axis, the scattered photon intensity follows the theoretically predicted $\mathrm{\cos^2\phi}$ dependence. The polarization angles (AOP) extracted via Equation~\ref{eq::polarization_fitting} consistently align with the beam spot's short axis at \SI{45}{\degree}.

Further analysis of secondary scattered photon intensities across all scanning points reveals the extracted AOP distribution in Fig.~\ref{fig::Polar_Direction} (a). Measurement positions (small circular markers) correspond to collimator displacements relative to the $\gamma$-ray beam centre. At the beam center (incident polarized $\gamma$-ray axis), AOP strictly follows the short axis direction at \SI{45}{\degree}. This result agrees with HI$\gamma$S findings~\cite{HIGS_Yan_Nature_Photonics}. Within proximal beam regions ($\le$\SI{2}{\mm} from axis), AOP remains uniformly distributed near \SI{45}{\degree}. However, in lower-intensity peripheral zones, AOP vectors exhibit tangential alignment relative to concentric circles.

Figure~\ref{fig::Polar_Direction} (b) compares the azimuthal distributions of incident $\gamma$-ray AOP from four scanning directions against Stokes parameter calculations (see S6). The measurements show good agreement with theoretical predictions. Notably, under relativistic conditions, scattered photons exhibit distributed AOP within the $1/\gamma$ angular range rather than maintaining a fixed polarization angle. This demonstrates that experiments using linearly polarized $\gamma$-rays from laser Compton sources require careful selection of polarization beam spot size to ensure measurement fidelity. 

\subsection{Asymmetry and DOP distribution}\label{subsec::asymmetry_and_polarization_degree}

\begin{figure*}[htb]
    \centering
    \includegraphics[width=\textwidth]{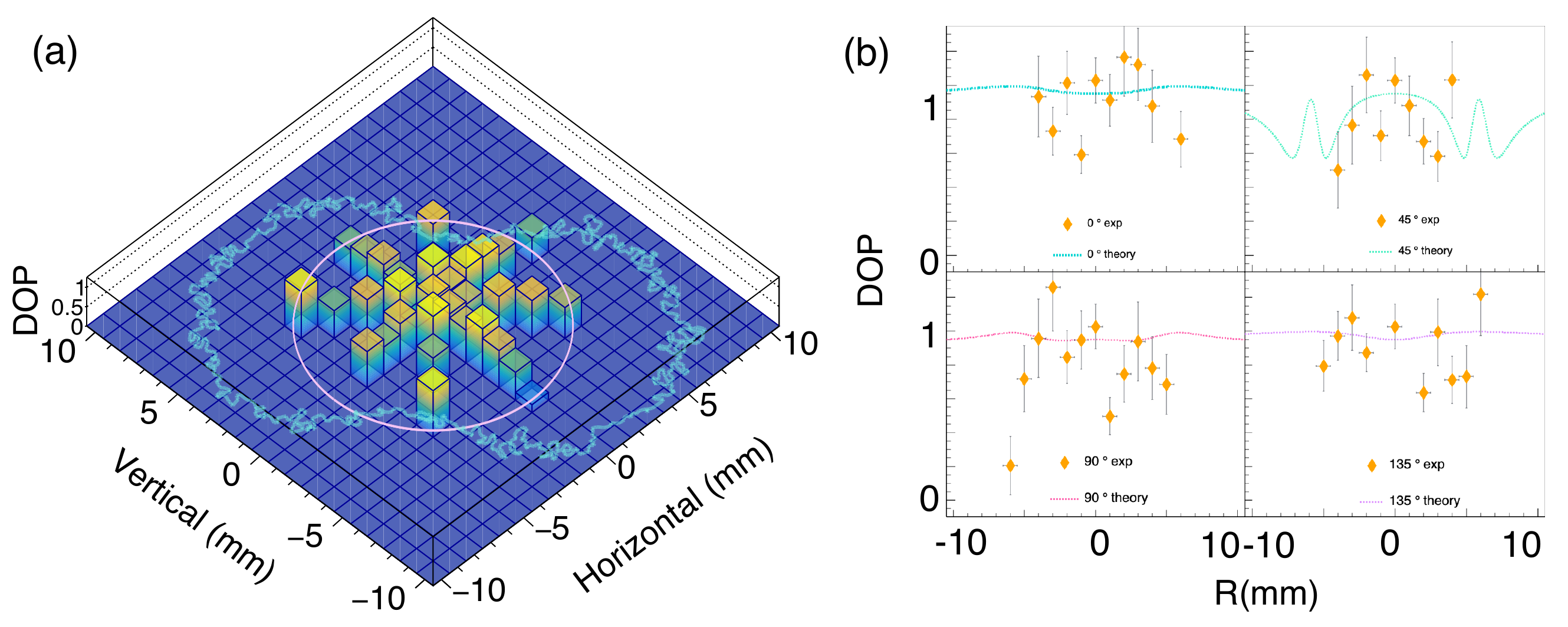}
    \caption{The experimental DOP distribution is shown in (a), where the blue line in the shape of peanut indicates $\gamma$-rays spot and the circle with pink represents the $1/\gamma$. A comparison between the experimental and calculation based on QED~\cite{WangXF_SLEGS_Polarization} is presented in (b).}
    \label{fig::Polaization}
\end{figure*}

The average DOP of the $\gamma$-ray beam within the \SI{1}{\mm} diameter collimator is obtained at all scanning points using Eq.~\ref{eq::polarization_degree}, and shown in Fig.~\ref{fig::Polaization} (a). The uncertainty of DOP mainly comes from statistical errors (less than 5.6\% in R0 data), efficiency correction errors (less than 3.2\%) and geometry correction errors(less than 0.3\%) (see S8).

Near the beam center, the measured DOP approaches 1.0, while it exhibits significant fluctuations in the peripheral regions. Fig.~\ref{fig::Polaization} (b) compares experimental and theoretical DOP values along four scanning directions, showing consistent agreement. 

Near-complete polarization transfer from laser photons to $\gamma$-rays is observed at the beam center (DOP $\approx$ 1.0) in  this case with the incident angle \SI{45}{\degree}. This demonstrates that \SI{45}{\degree} slant scattering provides an alternative pathway to high-DOP $\gamma$-rays beyond traditional \SI{180}{\degree} backward ICS. 

Reference~\cite{ChiZhijun_Thomson} extends the polarization property calculation from\SI{180}{\degree}  backward ICS to arbitrary slant geometries using a dipole radiation model, predicting geometry-independent polarization transfer while noting degraded DOP and bandwidth with increasing collection angle. Similarly, QED-based simulations in ~\cite{Filipescu_Monte_Carlo} confirm this conclusion. Our measured peripheral DOP exhibits analogous spatial variations. However, current experimental uncertainties prevent definitive quantitative validation of theoretical DOP distributions in peripheral regions, limiting complete characterization of polarization transfer efficiency in slant scattering geometries. In high-energy inverse Compton scattering (ICS) sources, the contribution of peripheral $\gamma$-rays becomes non-negligible due to collimator limitations and could introduce systematic errors to polarization-sensitive measurements. This underscores the need for higher-precision polarization measurements in slant ICS experiments. 

\section{CONCLUSION}\label{sec::conclusion}

Addressing the critical knowledge gap in polarization properties of photons scattered from relativistic electrons, we conducted the first systematic measurement of spatial polarization distribution for $\gamma$-rays generated via  \SI{45}{\degree} slant Inverse Compton Scattering between linearly polarized ($>93\%$) , \SI{0.117}{\eV}  ($\lambda$ = \SI{10.64}{\um}) \ce{CO_2} laser photons and \SI{3.5}{\GeV} relativistic electrons. Leveraging the fundamental principle that linearly polarized $\gamma$-rays induce asymmetric angular distributions in secondary scattered photons, we systematically measured intensity profiles, polarization angles (AOP), asymmetry parameters, and degree of polarization (DOP). The resulting linearly polarized $\gamma$-ray beam exhibits an asymmetric intensity profile, consistent with \SI{180}{\degree} backward ICS observations at HI$\gamma$S~\cite{HIGS_Yan_Nature_Photonics} and NewSUBARU~\cite{NewSUBARU} , while the spot is ``pinched" along the laser's AOP direction. At the beam center, AOP strictly aligns with the short-axis direction at \SI{45}{\degree} (matching HI$\gamma$S conclusions~\cite{HIGS_Yan_Nature_Photonics}), with DOP measuring $\approx$ 1.0 Near the beam axis, AOP remains uniformly distributed near \SI{45}{\degree}, while peripheral regions show tangential alignment relative to concentric circles and complex DOP variations.

These results demonstrate equivalent polarization transfer efficiency near the beam axis between \SI{45}{\degree} slant ICS and traditional \SI{180}{\degree} backward ICS, establishing slant scattering as a viable alternative for generating high-DOP $\gamma$-rays, significantly expanding the parameter space for polarized gamma sources. The uncertainties of current experiment limit precise DOP quantification in peripheral regions, precluding full characterization of polarization transfer efficiency in slant geometries. Future investigations are planned at the Shanghai Laser Electron Gamma Source (SLEGS) for enhanced polarization distribution measurements.

\bibliography{nsr_bib}
\end{document}